\documentstyle[twoside,epsf]{article}

\catcode`\@=11
\long\def\@makefntext#1{
\protect\noindent \hbox to 3.2pt {\hskip-.9pt  
$^{{\eightrm\@thefnmark}}$\hfil}#1\hfill}		%CAN BE USED 

\def\@makefnmark{\hbox to 0pt{$^{\@thefnmark}$\hss}}	%ORIGINAL 
	
\def\ps@myheadings{\let\@mkboth\@gobbletwo
\def\@oddhead{\hbox{}
\rightmark\hfil\eightrm\thepage}   
\def\@oddfoot{}\def\@evenhead{\eightrm\thepage\hfil
\leftmark\hbox{}}\def\@evenfoot{}
\def\sectionmark##1{}\def\subsectionmark##1{}}

{\Large
\oddsidemargin=\evensidemargin
\addtolength{\oddsidemargin}{-30pt}
\addtolength{\evensidemargin}{-30pt}

\newcounter{sectionc}\newcounter{subsectionc}\newcounter{subsubsectionc}
\renewcommand{\section}[1] {\vspace{12pt}\addtocounter{sectionc}{1} 
\setcounter{subsectionc}{0}\setcounter{subsubsectionc}{0}\noindent 
	{\tenbf\thesectionc. #1}\par\vspace{5pt}}
\renewcommand{\subsection}[1] {\vspace{12pt}\addtocounter{subsectionc}{1} 
	\setcounter{subsubsectionc}{0}\noindent 
	{\bf\thesectionc.\thesubsectionc. {\kern1pt \bfit #1}}\par\vspace{5pt}}
\renewcommand{\subsubsection}[1] {\vspace{12pt}\addtocounter{subsubsectionc}{1}
	\noindent{\tenrm\thesectionc.\thesubsectionc.\thesubsubsectionc.
	{\kern1pt \tenit #1}}\par\vspace{5pt}}
\newcommand{\nonumsection}[1] {\vspace{12pt}\noindent{\tenbf #1}
	\par\vspace{5pt}}

\topsep=0in\parsep=0in\itemsep=0in
\parindent=15pt

\newcommand{\textlineskip}{\baselineskip=13pt}

\def\abstracts#1#2#3{{
	\centering{\begin{minipage}{4.5in}\baselineskip=10pt\footnotesize
	\parindent=0pt #1\par 
	\parindent=15pt #2\par
	\parindent=15pt #3
	\end{minipage}}\par}} 

\renewenvironment{thebibliography}[1]
	{\frenchspacing
	 \ninerm\baselineskip=11pt
	 \begin{list}{\arabic{enumi}.}
        {\usecounter{enumi}\setlength{\parsep}{0pt}     
	 \setlength{\leftmargin 12.7pt}{\rightmargin 0pt} %FOR 1--9 ITEMS
         \setlength{\itemsep}{0pt} \settowidth
	{\labelwidth}{#1.}\sloppy}}{\end{list}}

\newcounter{itemlistc}
\newcounter{romanlistc}
\newcounter{alphlistc}
\newcounter{arabiclistc}

\def\@citex[#1]#2{\if@filesw\immediate\write\@auxout
	{\string\citation{#2}}\fi
\def\@citea{}\@cite{\@for\@citeb:=#2\do
	{\@citea\def\@citea{,}\@ifundefined
	{b@\@citeb}{{\bf ?}\@warning
	{Citation `\@citeb' on page \thepage \space undefined}}
	{\csname b@\@citeb\endcsname}}}{#1}}

\newif\if@cghi
\def\cite{\@cghitrue\@ifnextchar [{\@tempswatrue
	\@citex}{\@tempswafalse\@citex[]}}
\def\citelow{\@cghifalse\@ifnextchar [{\@tempswatrue
	\@citex}{\@tempswafalse\@citex[]}}
\def\@cite#1#2{{$\null^{#1}$\if@tempswa\typeout
	{IJCGA warning: optional citation argument 
	ignored: `#2'} \fi}}

\def\@refcitex[#1]#2{\if@filesw\immediate\write\@auxout
	{\string\citation{#2}}\fi
\def\@citea{}\@refcite{\@for\@citeb:=#2\do
	{\@citea\def\@citea{, }\@ifundefined
	{b@\@citeb}{{\bf ?}\@warning
	{Citation `\@citeb' on page \thepage \space undefined}}
	\hbox{\csname b@\@citeb\endcsname}}}{#1}}

\def\@refcite#1#2{{#1\if@tempswa\typeout
        {IJCGA warning: optional citation argument
	ignored: `#2'} \fi}}

\def\refcite{\@ifnextchar[{\@tempswatrue
	\@refcitex}{\@tempswafalse\@refcitex[]}}

\def\pmb#1{\setbox0=\hbox{#1}
	\kern-.025em\copy0\kern-\wd0
	\kern.05em\copy0\kern-\wd0
	\kern-.025em\raise.0433em\box0}

\def\fnt#1#2{\footnotetext{\kern-.3em
	{$^{\mbox{\scriptsize #1}}$}{#2}}}

\headsep=15pt   
   
\font\tenrm=cmr10
\font\tenit=cmti10 
\font\tenbf=cmbx10
\font\bfit=cmbxti10 at 10pt
\font\ninerm=cmr9

\font\eightrm=cmr8

\textwidth=5truein
\textheight=7.8truein

\def\qed{\hbox{${\vcenter{\vbox{			%HOLLOW SQUARE
   \hrule height 0.4pt\hbox{\vrule width 0.4pt height 6pt
   \kern5pt\vrule width 0.4pt}\hrule height 0.4pt}}}$}}

\begin{document}

%\runninghead{H.C. Rosu
%$\ldots$} {H.C. Rosu
%$\ldots$}

%Comment (HCR): produce fraza de mai sus la inceputul fiecarei pagini

\normalsize\textlineskip
\thispagestyle{empty}
\setcounter{page}{1}

%\copyrightheading{}                     %{Vol. 0, No.0 (1992) 000--000}

\vspace*{0.88truein}

%\fpage{1} %%%%%%%%%%%%%%%%%%%%%%%%%%%%%%%%%%%%%%%%%%%%%%%%%%%%%%%%%%%

\centerline{\bf MULTIPLE PARAMETER STRUCTURE OF  MIELNIK'S ISOSPECTRALITY}
\centerline{\bf IN UNBROKEN SUSYQM}
%\centerline{quant-ph/9904007 v5}
\vspace*{0.035truein}
%\centerline{\bf MANUSCRIPTS USING COMPUTER SOFTWARE\footnote{For
%the title, try not to use more than 3 lines. Typeset the title
%in 10 pt Times Roman, uppercase and boldface.}}
\vspace*{0.37truein}
\centerline{\footnotesize HARET C. ROSU}
\centerline{\footnotesize rosu@ifug3.ugto.mx}
%\footnote{Typeset names in
%10 pt Times Roman, uppercase. Use the footnote to indicate the
%present or permanent address of the author.}}
\vspace*{0.015truein}
\centerline{\footnotesize\it Instituto de F\'{\i}sica,
Universidad de Guanajuato, Apdo Postal E-143, Le\'on, Gto, Mexico}
\baselineskip=10pt
%\centerline{\footnotesize\it City, State ZIP/Zone,
%Country\footnote{State completely without abbreviations, the
%affiliation and mailing address, including country. Typeset in 8
%pt Times Italic.}}
\vspace*{10pt}
%\centerline{\footnotesize SECOND AUTHOR}
%\vspace*{0.015truein}
%\centerline{\footnotesize\it Group, Laboratory, Address}
%\baselineskip=10pt
%\centerline{\footnotesize\it City, State ZIP/Zone, Country}
%\vspace*{0.225truein}
%\publisher{(April 1, 1999)}{(April 1, 1999)}

%%%%%%%%%%%%%%%%%%%%%%%%%%%%%%%%%%%%%%%%%%%%%%%%%%%%%%%%%%%%%%%%%%%%%%%%
\vspace*{0.21truein}
\abstracts{
{\bf Abstract}.
Within unbroken SUSYQM and for zero factorization energy, I present an
iterative generalization of Mielnik's isospectral
method by employing a Schr\"odinger true zero mode in the first-step general 
Riccati
solution and imposing the physical condition of normalization at each
iterative step. This procedure leads to a well-defined multiple-parameter
structure within Mielnik's construction for both zero modes and potentials.
}{}{}
%%%%%%%%%%%%%%%%%%%%%%%%%%%%%%%%%%%%%%%%%%%%%%%%%%%%%%%%%%%%%%%%%%%%%%%%%%%

\vspace{2cm}

Int. J. Theor. Phys. {\bf 39}, 105-114 (January 2000)

%\vspace*{10pt}
%\keywords{The contents of the keywords}

\textlineskip                  %) USE THIS MEASUREMENT WHEN THERE IS
\vspace*{12pt}                 %) NO SECTION HEADING

\vspace*{1pt}\textlineskip	%) USE THIS MEASUREMENT WHEN THERE IS
%\section{General Appearance}    %) A SECTION HEADING
\vspace*{-0.5pt}
\noindent

%%%%%%%%%%%%%%%%%%%%%%%%%%%%%%%%%%%%%%%%%%%%%%
%PACS number(s):  98.80.Hw, 11.30.Pb

\noindent
%%%%%%%%%%%%%%%%%%%%%%%%%%%%%%%%%%%%%%%%%%%%%%%%%%%%%%%%%%%%%%%%%%%%%

%\newpage

%\pagebreak

%\textheight=7.8truein
%\setcounter{footnote}{0}
%\renewcommand{\thefootnote}{\alph{footnote}}

%\section{The Main Text}
\noindent

%$^{\ddagger}$
%{\it Institute of Gravitation and Space Sciences, P.O. Box MG-6,
%Magurele-Bucharest, Romania}

%\end{center}

%\bigskip
%\bigskip

%\begin{abstract}

%\end{center}
%{\em PACS}:  11.30.Pb\\
%{\em Keywords}: KW1, KW2

%\vskip 2cm

%%%%%%%%%%%%%%%%%%%%%%%%        THE PAPER       %%%%%%%%%%%%%%%%%%%%%%%%

\newpage
%{\Large
%\section{Introduction}

%\noindent
The supersymmetric procedures are an interesting
and fruitful extension of (one-dimensional) quantum mechanics. For recent
reviews see [\refcite{susy}]. These techniques are, essentially,
factorizations of one-dimensional Schr\"odinger operators, first
discussed in the supersymmetric context
by Witten in 1981 [\refcite{w81}], and well known in the mathematical
literature in the
broader sense of Darboux covariance of Schroedinger equations
[\refcite{d82}].

In 1984, Mielnik [\refcite{mn}] introduced a different factorization of the
quantum harmonic oscillator based on the general Riccati solution. As a result,
Mielnik obtained a one-parameter
family of potentials with {\em exactly} the same spectrum as that of the
harmonic oscillator. However, even though in the same year Nieto discussed the
connection
of such a factorization with the inverse scattering approach, and Fern\'andez
applied it to the hydrogen atom case, Mielnik's result remained a curiosity
for a decade during which only a few authors paid attention to it.
On the other hand, constructing families of strictly isospectral potentials is
an important possibility with many potential
applications in physics [\refcite{susy}]. This explains the recent surge
of interest in this supersymmetric issue [\refcite{rosu}].
My goal in this work is to give a multiple-parameter generalization of
Mielnik's procedure based on the ground-state function of any soluble
one-dimensional quantum mechanical problem. This is just
a form of Crum's iterations, i.e., repeated Darboux transformations.
Some work along this line has already been done by
Keung {\em et al.} [\refcite{k}], who performed an iterative construction for
the reflectionless, solitonic, ${\rm sech}$ potentials and the attractive
Coulomb potential presenting
relevant plots as well. However, they first go $n$ steps away from a given
ground state and only afterwards perform the $n$ steps backwards.
On the other hand, Pappademos {\em et al.} [\refcite{psp}],
working in the continuum part of the spectrum,
got one- and two-parameter supersymmetric families of potentials strictly
isospectral with respect to the half-line free particle and Coulomb potentials
and focused on the supersymmetric bound states in the continuum. Their procedure
is closer to the method I will present in the following.
For more recent works see
Bagrov and Samsonov [\refcite{bs}], Fern\'andez {\em et al} [\refcite{fvm}],
Junker and Roy [\refcite{jr}], and Rosas-Ortiz [\refcite{roo}].

In the following, I first briefly recall the mathematical background of
Mielnik's method and next pass to a simple multiple-parameter
generalization for the particular but physically relevant zero-mode case.

%{\bf 2}. -
%%%%%%%%%%%%%
I begin with the ``fermionic" Riccati (FR) equation
$y^{'}=-y^2+V_{1}(x)$ [the ``bosonic" one being $y^{'}=y^2+V_{0}(x)$] for which
I suppose to know a particular solution $y_{0}$. Notice also that I do not
put any free constant in the Riccati equations, that is, I work at zero
factorization energy. Let us seek the general solution in the form
$y_1=w_1+y_{0}$.
By substituting $y_{1}$ in the FR equation  one gets the Bernoulli equation
$-w_{1}^{'}=w_{1}^2+(2y_{0})w_{1}$. Furthermore,
using $w_2=1/w_1$, we obtain the simple first-order linear differential equation
$w_{2}^{'}-(2y_{0})w_{2}-1=0$, which can be solved
by employing the
integration factor $F_{0}(x)=e^{-\int _{c}^{x}2y_{0}}$, leading
to the solution
$w_2(x)=(\lambda +\int _{c}^{x}F_{0}(z)dz)/F_{0}(x)$,
where $\lambda$ occurs as an integration constant. In applications the lower
limit $c$ is either $-\infty$ or 0 depending on whether one deals with full-line
or half-line problems, respectively. In the latter case, $\lambda$ is
restricted to be a positive number.
Thus, the general FR solution reads
%%%%%%%%%%%%%%%%%%%%%%%%%%%%%%%%%%%%%%%
$$
y_{1}=y_{0}+
\frac{e^{-\int _{c}^{x}2y_{0}}}{\lambda +\int _{c}^{x}
e^{-\int _{c}^{z}2y_{0}}}\equiv
y_{0}+\frac{F_{0}}{\lambda  +\int _{c}^{x}(F_{0})}=y_{0}+
D\ln\left(\lambda  +\int _{c}^{x}(F_{0})\right)~,
\eqno(1)
$$
%%%%%%%%%%%%%%%%%%%%%%%%%%%%%%%%%%%%%%%%%   1
where $D=\frac{d}{dx}$.
%For the usual nonrelativistic one-dimensional quantum mechanics
%$b=0$ and $a=-1$, i.e., the integration factor is
%$f_{qm}=e^{-2\int ^{x}y_{0}}$.
It is easy to reach the conclusion that the particular FR
solution $y_{0}$ corresponds to
Witten's superpotential [\refcite{w81}], %in supersymmetric quantum mechanics
%\cite{SQM},
while the general FR solution
$y_1=y_{0}+\frac{F_{0}}{\lambda +\int _{c}^{x}F_{0}}$ is of
Mielnik type [\refcite{mn}]. This is especially clear when one is able to
identify $f_{0}=F_{0}^{1/2}$ with the quantum mechanical
ground-state wavefunction $u_{0}$ of the problem at hand.
This requires suitable asymptotic behaviour of the Riccati solution $y_{0}$
and applying the normalization condition to $f_{0}$,
turning it into a true zero mode. As is well known, this case corresponds to
the so-called unbroken SUSYQM, which will be assumed to hold henceforth.
Moreover, $-2y_{0}^{'}$
($\equiv -2\frac{d^2}{dx^2}\ln F_{0}^{1/2}$)
is the Darboux transform contribution to the initial
Schr\"odinger potential, i.e., $V_1=V_0-2y_{0}^{'}$. Also, the modes
%%%%%%%%%%%%%%%%%%%%%%
$$
%\begin{equation} \label{2}
u _{\lambda}(x)=
\frac{F_{0}^{1/2}}{\lambda +\int _{c}^{x} F_{0}}=\frac{u _{0}}{\lambda +
\int _{c}^{x}u_{0}^2}
\eqno(2)
%\end{equation}
$$
%%%%%%%%%%%%%%%%%%%%              2
can be normalized and therefore considered as ground-state wavefunctions
of the bosonic family of potentials
corresponding to Mielnik's parametric superpotential.
The one-parameter true zero modes read
%%%%%%%%%
$$
v _{\lambda}(x)=
\frac{\sqrt{\Lambda}F_{0}^{1/2}}{\lambda +\int _{c}^{x} F_{0}}
=\frac{\sqrt{\Lambda}u _{0}}{\lambda +
\int _{c}^{x}u_{0}^2}~,
\eqno(3)
$$
%%%%%%%%%%  3
where $\sqrt{\Lambda}=\sqrt{\lambda(\lambda +1)}$ is the normalization
constant.
 
Moreover,
$-2y_{1}^{'}$ can be thought of as the general Darboux transform contribution
to the initial potential generating the bosonic strictly isospectral
family, which reads
%%%%%%%%%%%%%%%%%%%
%\begin{equation} \label{4}
$$
V_{\lambda}^{M}=
 V_{0}(x)-2\frac{d^2}{dx^2}\ln \left(\lambda +\int ^{x}u^{2}_{0}\right)
=V_{0}(x)-\frac{4u_{0}u^{'}_{0}}{\lambda +\int^{x}u^2_{0}}
+\frac{2u^4_{0}}{(\lambda +\int^{x}u^2_{0})^2}~.
%\end{equation}
\eqno(4)
$$
%%%%%%%%%%%%%%%%%%%%%%%%           4
This family of potentials can be seen as a continuous deformation of the
original potential, because the latter is included in the infinite limit of
the deforming parameter $\lambda$ and $v_{\pm\infty}=u_{0}$ as well.
%All these relationships are supplementary material to the core of
%the ``entanglement" between Riccati and Schroedinger equations,
%which has been recently emphasized by Haley \cite{h97}.
In more intuitive terms, Mielnik's method based on an intial Schroedinger
true zero mode may be called
a double Darboux technique of deleting followed
by reinstating a nodeless ground-state wavefunction
$u_0(x)$ of a potential $V_{0}(x)$ by
means of which one can generate a one-parameter family of isospectral
potentials
$V_{\lambda}(x)$, where $\lambda$ is a labeling,
real parameter of each member potential in the set.

One can go on with one of the strictly isospectral bosonic
zero modes $u_{\lambda _{1}}=\frac{u_0}{\lambda _{1}+\int _{c}^{x} u_{0}^{2}}$
(i.e., by choosing $\lambda =\lambda _{1}$)
and repeat the strictly isospectral construction, getting a
new two-parameter %(of which only the second one is a free parameter)
zero mode
%%%%%%%%%%%%%%%%%%%%%%
$$u _{\lambda _1,\lambda _{2}}=\frac{u_{\lambda _1}}{\lambda _2
+\int _{c}^{x} u_{\lambda _1}^{2}}=\frac{u_{0}}{(\lambda _1+\int _{c}^{x}
u_{0}^{2})
(\lambda _2 +\int _{c}^{x}u_{\lambda _1}^{2})}~.
\eqno(5)
$$
%%%%%%%%%%%%%%%%%%%%%%%%%%%%%%  5
The two-parameter true zero modes read
%%%%%%%%%
$$
v_{\lambda _1,\lambda _{2}}=
\frac{\sqrt{\Lambda _1\Lambda _2}
u_{0}}{(\lambda _1+\int ^{x}u_{0}^{2})
(\lambda _2 +\int ^{x}v_{\lambda _1}^{2})}~.
\eqno(6)
$$
%%%%%%%%%%%%%%%  6
The resulting
two-parameter family of strictly isospectral potentials will be
%%%%%%%%%%%%%%%%%
$$
V_{\lambda _{1}, \lambda _{2}}=
V_{0}-2\frac{d^2}{dx^2}\ln\Big[\left(\lambda _{1}+\int ^{x}u_{0}^{2}\right)
\left(\lambda _2 +\int ^{x}v_{\lambda _1}^{2}\right)\Big]~.
%V_{\lambda _{1}}(x)-\frac{4u_2u^{'}_{2}}{\lambda _{2} +\int^{x}u_{2}^2}
%+\frac{2u_{2}^4}{(\lambda _{2} +\int^{x}u_{2}^2)^2}~.
\eqno(7)
$$
%%%%%%%%%%%%%%%%%%%%%%%%   7
At the $i$th -parameter level, one will have
%%%%%%%%%%%%%%%%%%%
$$
V_{\lambda _{1}, \lambda _{2},... \lambda _{i}}=
V_{0}-2\frac{d^2}{dx^2}\ln\Big[\left(\lambda _{1}+\int ^{x}u_{0}^{2}\right)
\left(\lambda _2 +\int ^{x}v_{\lambda _1}^{2}\right)...
\left(\lambda _{i}+\int^{x}v_{\lambda _{1}...\lambda _{i-1}}^{2}\right)\Big]
\eqno(8)
$$
%%%%%%%%%%%%%%%%%%%%%%%%%%%%%%%%%%%%%%%  8
and
%%%%%%%%%%%%%
$$
v_{\lambda _{1}...\lambda _{i}}=
\frac{\sqrt{\Lambda _1 \Lambda _2...\Lambda _{i}}
u_{0}}{(\lambda _{1}+\int ^{x}u_{0}^{2})...(\lambda _{i}+
\int ^{x}v_{\lambda _{1}...\lambda _{i-1}}^{2})}~.
\eqno(9)
$$
%%%%%%%%%%%%%%%%%%%%%%%%%%%%%%  9

Explicit formulas for the parametric zero modes can be obtained if
one uses a notation based on the integration factor
$\int _{c}^{x}F_{0}={\cal F}(x)-{\cal F}(c)=\Delta _{x}{\cal F}$. Then
%%%%%%%%%
$$
v_{\lambda _1}(x)=
\frac{\sqrt{\Lambda _1}u_0}{
\lambda _1 +\Delta _{x}{\cal F}}~.
\eqno(10)
$$
%%%%%%%%
Next, one can calculate
$$
\int _{c}^{x}\frac{u_{0}^{2}}{(\lambda +\int _{c}^{x}
u_{0}^{2})^2}=\int _{c}^{x}
\frac{{\cal F}^{'}dx}{(\lambda -{\cal F}(c)+{\cal F}(x))^{2}}
=\int _{{\cal F}(c)}^{{\cal F}
(x)}\frac{dz}{(\lambda -{\cal F}(c)+z)^2}=
$$
$$
=\frac{1}{\lambda}
-\frac{1}{\lambda  +\Delta _{x}{\cal F}}=\frac{\Delta _{x}{\cal F}}
{\lambda(\lambda +\Delta _{x}{\cal F})}~.
\eqno(11)
$$
%%%%%%%%%%
Thus
%%%%%%%
$$
v_{\lambda _1,\lambda _2}(x)=\frac{\sqrt{\Lambda _1\Lambda _2}u_0}{
\lambda _1 \lambda _2+(\lambda _1+\lambda _2+1)\Delta _{x}{\cal F}}~.
\eqno(12)
$$
%%%%%%%%%%%%%%%
At the next step one gets
%%%%%%%%%%%%
$$
v_{\lambda _1,\lambda _2, \lambda _3}(x)=
\frac{\sqrt{\Lambda _1\Lambda _2\Lambda _3}u_0}{
\lambda _1 \lambda _2 \lambda _3+
(\lambda _1\lambda _2+\lambda _2\lambda _3+\lambda _3\lambda _1+
\lambda _1+\lambda _2+\lambda _3 +1)\Delta _{x}{\cal F}}
\eqno(13)
$$
%%%%%%%%%%%%
and the general formula at the $i$ level can be written down in the form
%%%%%%%%%%%
$$
v_{\lambda _1,\lambda _2,..., \lambda _{i}}(x)=
\frac{\sqrt{\Lambda _1....\Lambda _{i}}u_{0}}{C_{1}^{(i)}+
C_{2}^{(i)}\Delta _{x}{\cal F}}~,
\eqno(14)
$$
%%%%%%%%%%
where the first coefficient in the denominator is the product of all
parameters, whereas the second coefficient is just the sum over all the rest
of lower order Viete-type expressions in the parameters.
By the same token, one can write a general formula for the strictly
isospectral potentials
%%%%%%%%%%
$$
V_{\lambda _1,\lambda _2,...,\lambda _{i}}=V_{0}-
2D^2\ln[C_{1}^{(i)}+C_{2}^{(i)}\Delta _{x}{\cal F}]=
V_{0}-\frac{4C_{2}^{(i)}u_{0}u_{0}^{'}}{C_{1}^{(i)}+
C_{2}^{(i)}\Delta _{x}{\cal F}}+\frac{2(C_{2}^{(i)})^2u_{0}^{4}}{
(C_{1}^{(i)}+C_{2}^{(i)}\Delta _{x}{\cal F})^2}~,
\eqno(15)
$$
%%%%%%%%%%%%%
which may be considered as the generalization of the furthest right-hand side
of Eq.~(4) and
represents a simple generalization of Mielnik's one-parameter potentials.
%The singularities in Eq.~(14) and (15) are avoided if
%$|C_{1}^{(i)}+C_{2}^{(i)}\Delta {\cal F}|\neq 0$.
%%%%%%%%%%%%%

Since $\Lambda _1...\Lambda_{i}=C_{1}^{(i)}(C_{1}^{(i)}+C_{2}^{(i)})$, one
might
think that there is nothing new in (14) and (15) with respect to a common
Mielnik solution with an effective parameter $\lambda _{eff}^{(i)}
=C_{1}^{(i)}/C_{2}^{(i)}$. However, I will argue that by performing such an
equivalence one loses a certain type of information.
This information is a consequence of the symmetry of
(14) and (15) in the space of parameters. One can see that the subindices of
any pair of parameters can be interchanged without affecting the formulas.
Thus, each $\lambda$ parameter can be varied independently of the others,
making it possible to put questions related to the following type of situation.
Suppose we construct two Mielnik potentials corresponding to $\lambda _1$
and $\lambda _2$ and ask what is the potential bearing true zero modes that
for $\lambda _1\rightarrow \pm\infty$ goes to the Mielnik case
for $\lambda _2$,
whereas for $\lambda _2\rightarrow \pm\infty$ it goes to the Mielnik case for
$\lambda _1$. The answer is provided by the construction of this work and
corresponds to the particular case $V_{\lambda _1,\lambda _2}$ bearing the
true zero modes $v_{\lambda _1,\lambda _2}$. Indeed, as one can easily check,
$v_{\lambda _1,\pm\infty}=v_{\lambda _1}$ and
$V_{\lambda _1,\pm\infty}=V_{\lambda _1}^{M}$,
whereas $v_{\pm\infty, \lambda _2}
=v_{\lambda _2}$ and $V_{\pm\infty,\lambda _2}=V_{\lambda _2}^{M}$. In the
general case, one starts with a set of $i$ Mielnik potentials corresponding
to $i$ fixed values of Mielnik's parameter and asks the same question, this
time for the set of $i$ asymptotic limits. The answer is given by
(14) and (15) and cannot be provided if one works with only one effective
parameter unless its multiple-parameter value found above is used.

Another interesting remark is the one-to-one relationship between any
polynomial equation $a_0x^{i}+a_{1}x^{i-1}+...+a_{i}=0$ and the present
iterative construction. If we consider the $\lambda$ parameters as the
zeros of such arbitrary polynomials, we can write (14) as
$$
v_{\lambda _1,\lambda _2,..., \lambda _{i}}=
\frac{\sqrt{(-1)^{i}a_{i}(\sum _{0}^{i}(-1)^{i}a_{i})}u_{0}}{(-1)^{i}a_i+
(\sum _{0}^{i-1}(-1)^{i}a_i)\Delta _{x}{\cal F}}~,
\eqno(16)
$$
and in (15) one can substitute the same type of denominator as in (16).
There is only one constraint on the employed polynomials, which one
should impose in order to avoid possible singularities.
Usually the integral $\Delta _{x}{\cal F}$
in the denominators of (15) and (16) is of the kink type, i.e., it may be
written in the form $\alpha +\beta K(x)$, where the function $K(x)$ has the
kink behavior,
taking values between -1 and +1 and $\alpha$ and $\beta$ are some constants, of
which $\alpha$ may be zero. Then the allowed intervals for the
effective parameter are $\lambda _{eff}^{(i)}>\beta -\alpha$ and
$\lambda _{eff}^{(i)}<-(\beta +\alpha)$. When $\alpha =0$ one gets
$|\lambda _{eff}^{(i)}|> \beta$.
%For example, for the case of the
%isotropic oscillator $\alpha =0$ and $\beta =1/2$.
%The implications of this type of mapping between polynomials and `zero modes'
%is beyond the goals of this work.
% (I call `zero modes' even those
%corresponding to possible complex values of the parameters [\refcite{bey}]).

It is worthwhile to mention that the previous iteration process can be
understood most easily from
the Riccati equation standpoint as follows. To get, for example,
the two-parameter zero
mode, one should start again with the FR equation $y^{'}=-y^{2}+V_{1}(x)$
and take as the known particular solution $y_{p}^{(1)}=y_{0}+y_{\lambda _{1}}$,
where $y_{\lambda _{1}}=\frac{F_{0}}{\lambda _{1}  +\int _{c}^{x}(F_{0})}$.
The intermediate Bernoulli equation will
be $-w_{1}^{'}=w_{1}^{2}+2y_{p}^{(1)}w_{1}$. This is turned into a
first-order
differential equation by the inverse function method. The integration factor
of the latter is
$F_{\lambda _1}={\rm exp}^{(-\int _{c}^{x}2y_{p}^{(1)})}$ and the solution for 
the first-order differential equation can be written $w_2=(\lambda _2+\int _{c}^{x}
F_{\lambda _1}dz)/F_{\lambda _1}$. From this presentation it is clear how one
should proceed for an arbitrary step. Also, the logarithmic derivative notation
in Eq. (1) is equally convenient to have a clear image of the iteration
process. Thus, one can generate hierarchies of parametric
Schr\"odinger zero modes of any desired order by means of the Riccati
connection.

The parametric
normalization deletes the interval $[-1,0]$ from the parameter space
of $\lambda _{eff}^{(i)}$.
At the $-1$ limit, one can make a connection with the Abraham-Moses
isospectral technique [\refcite{am}], whereas at the $0$ limit the connection
can be done
with another isospectral construction developed by Pursey [\refcite{p}].
This connection is only from the point of view of the potentials; the
zero modes as worked out here just disappear.
%Moreover, since the strictly isospectral supersymmetry obviously may
%introduce singularities in both wavefunction and potential, usually the
%active authors in the field discard those values of the deformation
%parameter for which those singularities occur. If
%$|\lambda _{{\rm sing}}|\geq 1$ the excluded interval in the parameter space
%is $[-\lambda _{{\rm sing}}, \lambda _{{\rm sing}}]$, and therefore the
%connections with the Abraham-Moses and Pursey methods are lost in this case.
%Moreover, because of the singularity-free constraint
%on $\lambda _{eff}^{(i)}$, usually at least one of the limits is lost,
%as for example in the harmonic oscillator case where the excluded interval
%is $[-\frac{\sqrt{\pi}}{2},\frac{\sqrt{\pi}}{2}]$ implying the loss of the
%Pursey limit.

%\section{Conclusion}

In conclusion, I have shown explicitly the way Crum's iteration works when
the general Riccati solutions (general superpotentials) at zero factorization
energy are based on the
corresponding Schr\"odinger ground-state wavefunctions, obtaining
general formulas for this simple `generalization' of Mielnik's
one-parameter SUSYQM isospectrality. Plots of the two-parameter
formulas for the harmonic oscillator case are presented 
in Fig. 1-5.
One may consider the
results of this work as pointing to an interesting hierarchical
structure within the general Riccati solution
produced by a particular type of repeated Darboux transformations
when the normalization condition of quantum mechanics is taken care of at each
iterative step.

%%%%%%%%%%%%%%%%%%%%%%%%%%%%%%%%%%%%%%%%%%%%%%%%%%%%%%%%%%%%%%%%%%%%%%
%\begin{thebibliography} {99}
%\bibitem[*]{byline} Electronic address: rosu@ifug.ugto.mx

%\end{thebibliography}
%%%%%%%%%%%%%%%%%%%%%%%%%%%%%%%%%%%%%%%%%%%%%%%%%%%%%%%%%%%%%%%%%%%%%%%%%

%\bigskip
%\newpage
%{\bf Figure Caption}
%\bigskip

%Fig. 1  Members of the

%Fig. 2 Squares of the

%%%%%%%%%%%%%%%%%%%%%%%%%%%%%%%%%%%%%%%%%%%%%%%%%%%%%%%%%%%%%%%%%%%%%%%%%
 \newpage

\nonumsection{Acknowledgments}
\noindent
%\section*{Acknowledgment}
This work was partially supported by the CONACyT Project 458100-5-25844E.
I wish to thank D.J. Fern\'andez and B. Mielnik for very useful
correspondence.

%This section should come before the References. Funding
%information may also be included here.

%\newpage
\nonumsection{References}
%\noindent
%References are to be listed in the order cited in the text. Use
%the style shown in the following examples. For journal names,
%use the standard abbreviations. Typeset references in 9 pt Times
%Roman.

%\newpage
%%%%%%%%%%%%%%%%%%%%%%%%%%%%%%%%%%%%%%%%%%%%%%%%%%%%%%%%%%%%%%%%%%%%%%%%%

\newpage

%{\bf Figure captions}

%\bigskip

\centerline{
\epsfxsize=290pt
\epsfbox{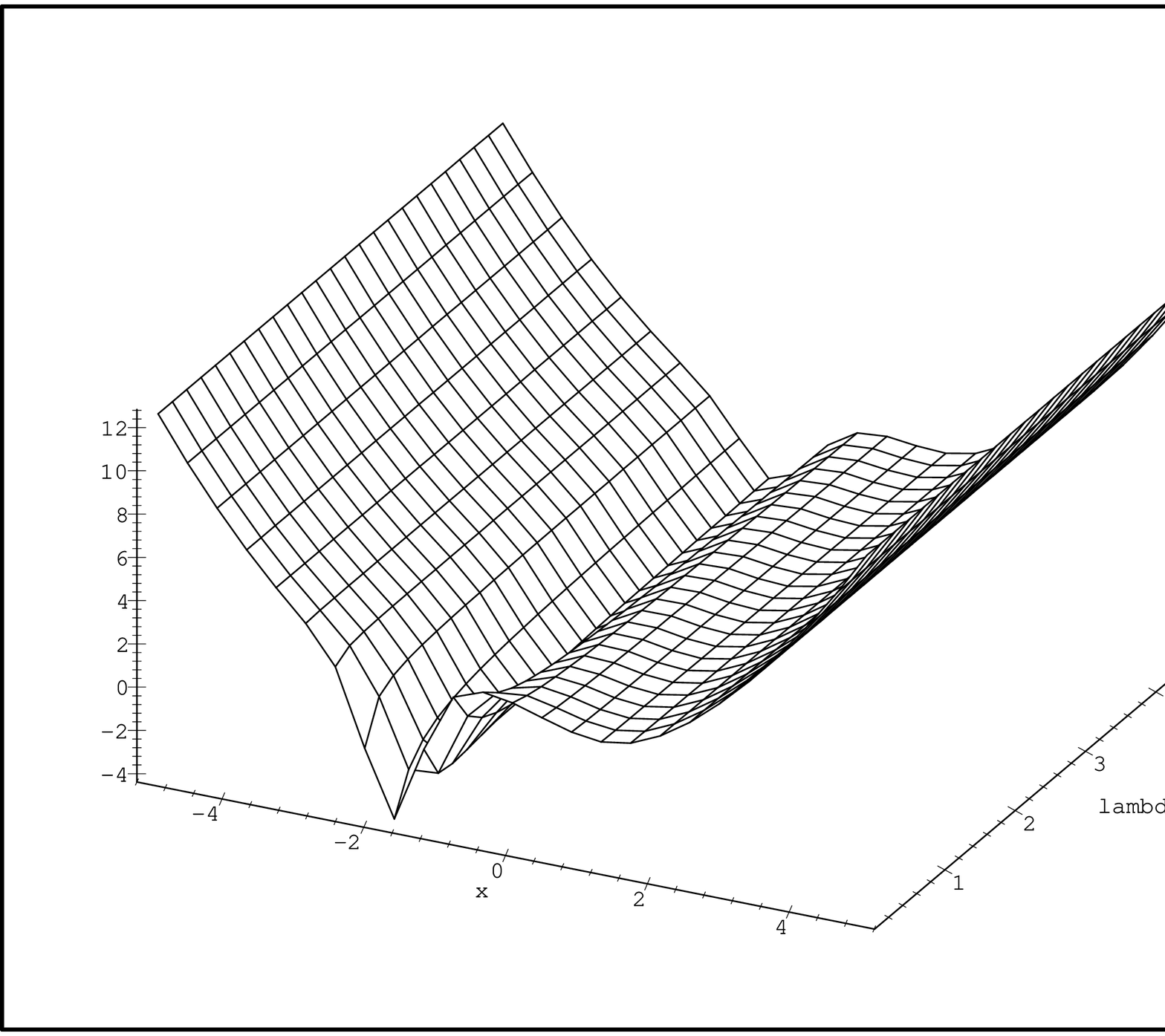}}
\vskip 4ex
\begin{center}
{\small{Fig. 1}\\
Two-parameter strictly isospectral harmonic oscillator potentials
($\hbar=m=\omega=1$), for
fixed $\lambda _2=0.2$ and $\lambda _1\in[0.1,5]$. They are identical to
Mielnik harmonic oscillator potentials with $\lambda _{eff}^{(2)}\in 
[0.0154,0.1613]$.}
\end{center}

%\noindent
%Fig. 1: Two-parameter strictly isospectral harmonic oscillator potentials
%fixed $\lambda _2=0.2$ and $\lambda _1\in[0.1,5]$. They are identical to
%Mielnik harmonic oscillator potentials with $\lambda _{eff}\in [0.0154,0.1613]$.

%\bigskip

\newpage

\centerline{
\epsfxsize=280pt
\epsfbox{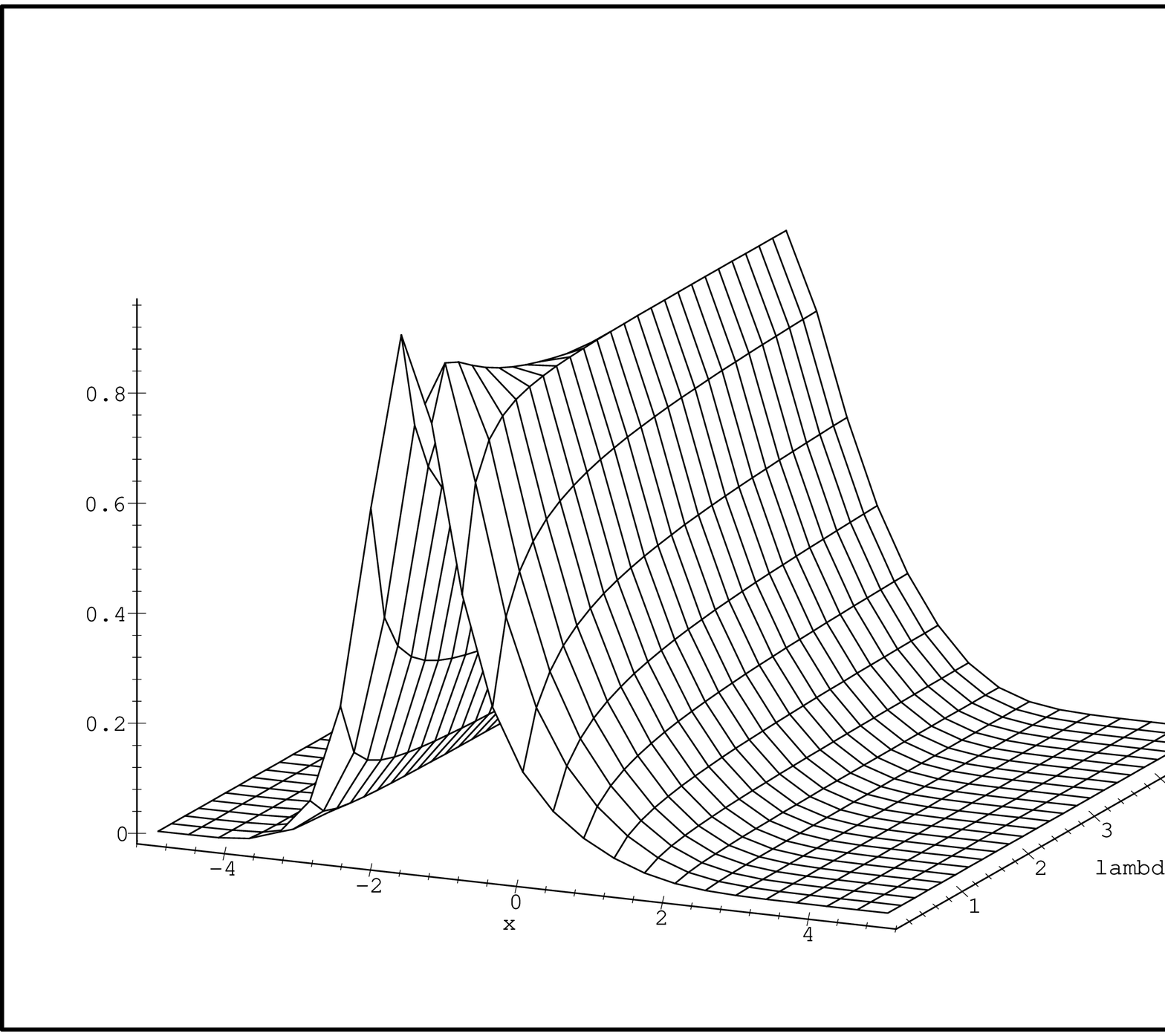}}
\vskip 4ex
\begin{center}
{\small{Fig. 2}\\
The corresponding true zero modes.}
\end{center}

%\noindent
%Fig. 2: The true zero modes corresponding to the potentials of Fig. 1.

%\bigskip

\newpage

\centerline{
\epsfxsize=290pt
\epsfbox{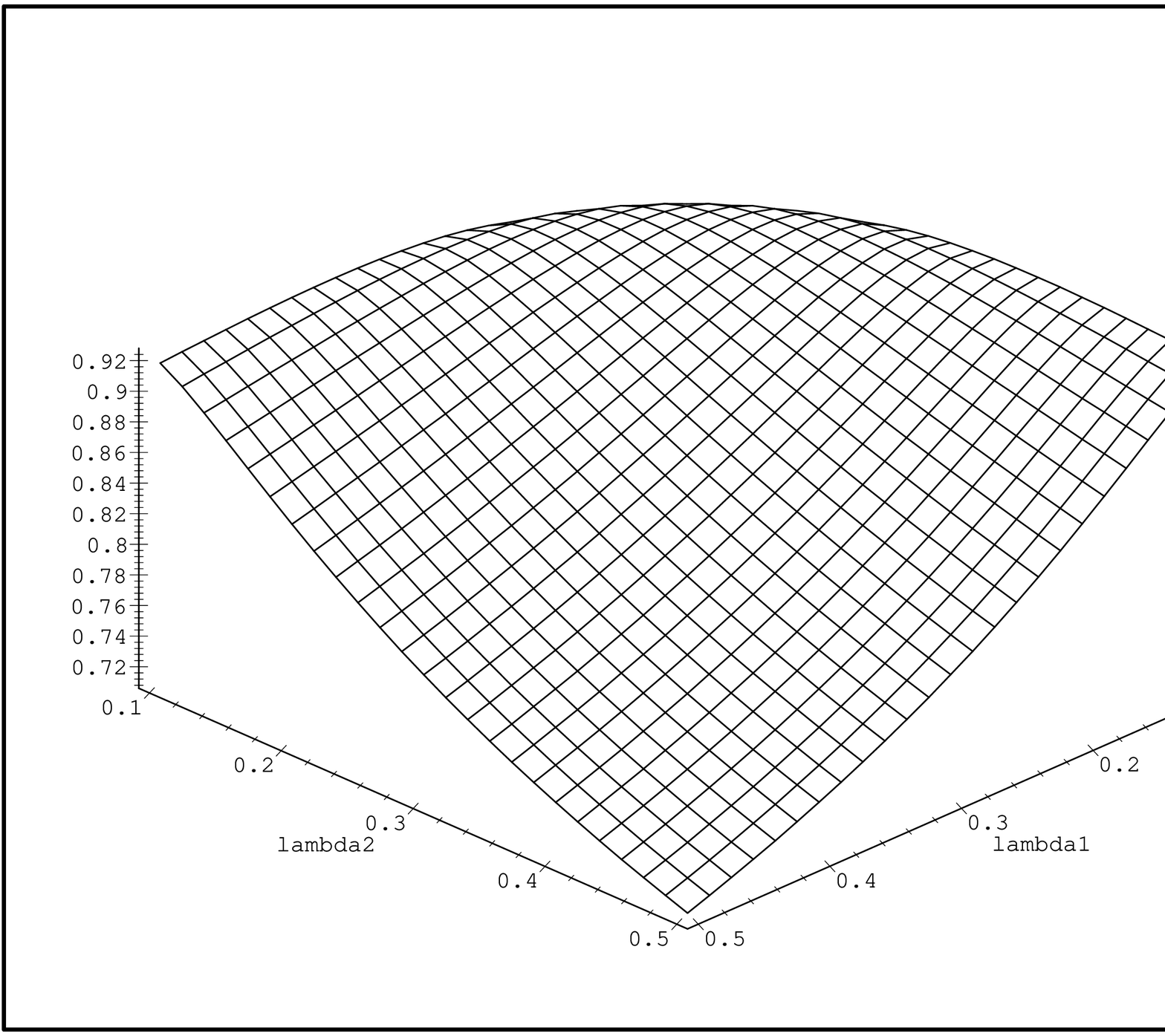}}
\vskip 4ex
\begin{center}
{\small{Fig. 3}\\
The two-parameter true zero modes at fixed $x=-1.4$ as a function of the
two parameters.}
\end{center}

\newpage

\centerline{
\epsfxsize=280pt
\epsfbox{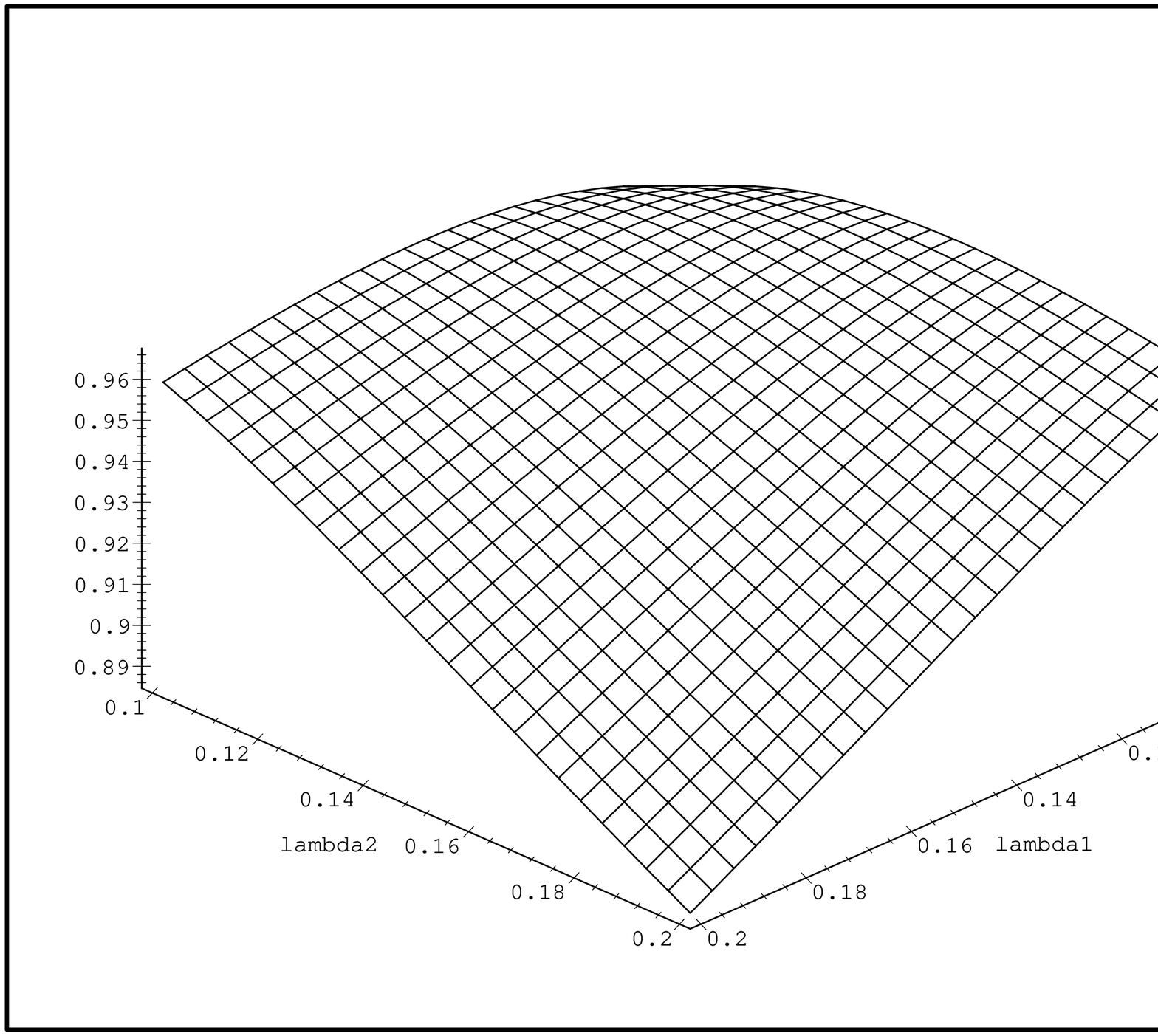}}
\vskip 4ex
\begin{center}
{\small{Fig. 4}\\
 Same modes as in Fig. 3 at fixed $x=-1.6$.}
\end{center}

\newpage

\centerline{
\epsfxsize=280pt
\epsfbox{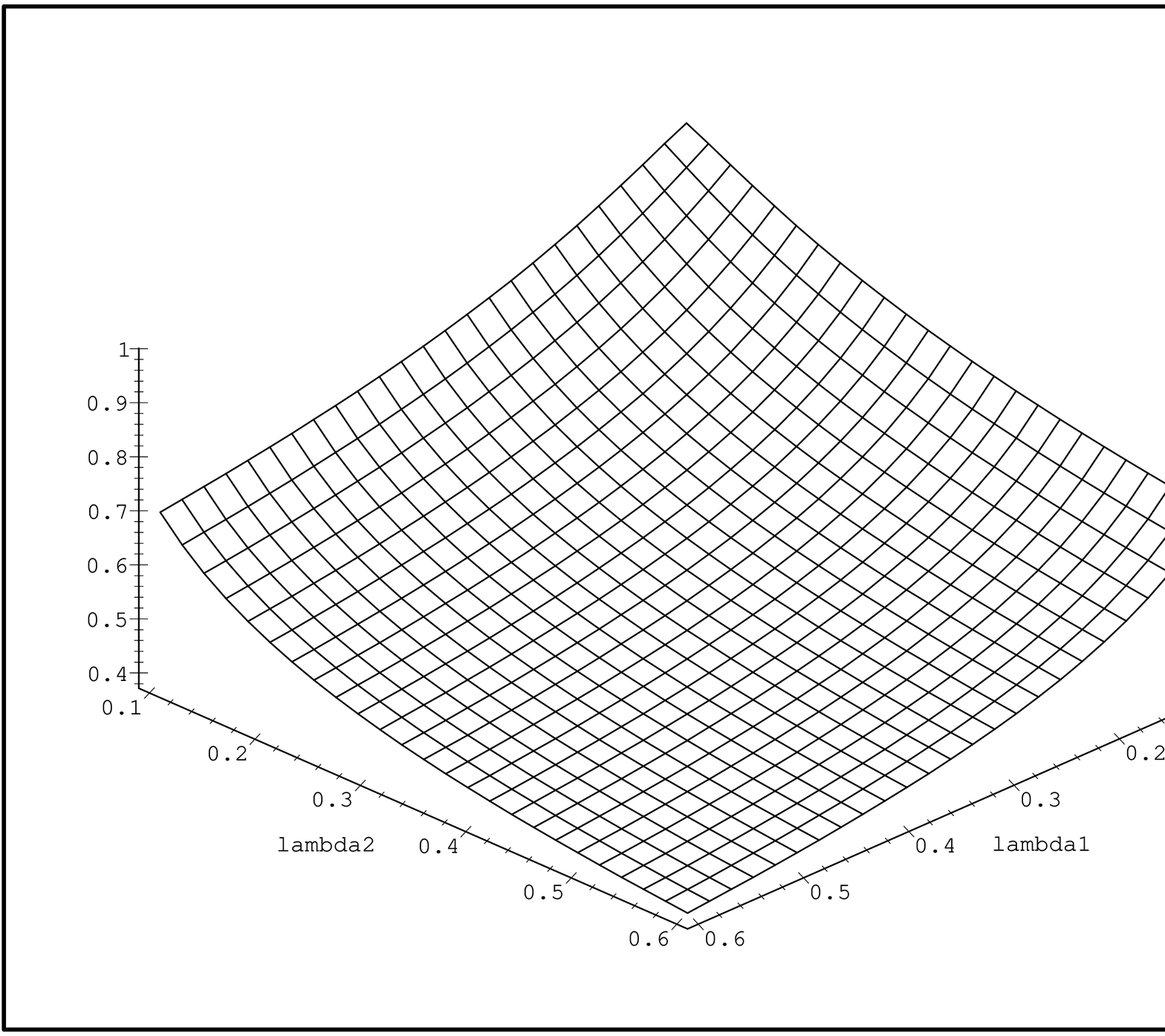}}
\vskip 4ex
\begin{center}
{\small{Fig. 5}\\
 Same modes as in Fig. 3 for fixed $x=-1.8$.}
\end{center}

%\appendix

%\noindent
%Appendices should be used only when absolutely necessary. They
%should come after the References. If there is more than one
%appendix, number them alphabetically. Number displayed equations
%occurring in the Appendix in this way, e.g.~(\ref{that}), (A.2),
%etc.
%\begin{equation}
%\mu(n, t) = {\sum^\infty_{i=1} 1(d_i < t, N(d_i) = n) \over
%\int^t_{\sigma=0} 1(N(\sigma) = n)d\sigma}\,. \label{that}
%\end{equation}
}
\end{document}